\documentclass[cameraready]{Interspeech}

\title{Physics-Informed Neural Operator for Speech Production Analysis}

\author[affiliation={1}, orcid=0000-0001-5995-3744, correspondingauthor]{Kazuya}{Yokota}
\author[affiliation={2}, orcid=0009-0004-9283-5511]{Xinmeng}{Luan}
\author[affiliation={3}, orcid=0000-0002-1541-1529]{Debasish Ray}{Mohapatra}
\author[affiliation={2}]{Gary}{Scavone}
\author[affiliation={3}, orcid=0000-0001-9279-9021]{Sidney}{Fels}

\address{
    $^1$ Department of Mechanical Engineering, Nagaoka University of Technology, Japan \\
    $^2$ Schulich School of Music, McGill University, Canada \\
    $^3$ Department of Electrical and Computer Engineering, University of British Columbia, Canada
}

\email{yokokazu@vos.nagaokaut.ac.jp, xinmeng.luan@mail.mcgill.ca, debasishray@ece.ubc.ca, gary.scavone@mcgill.ca, ssfels@ece.ubc.ca}

\keywords{speech production, vocal folds, physics-informed neural networks, neural operators, DeepONet}

\usepackage{comment}

\begin{document}

\maketitle

\begin{abstract}
Physics-informed neural operators (PINOs) have recently gained attention as fast numerical simulators with potential for solving inverse problems. This study proposes the first PINO-based method for speech production analysis. The model learns the governing one-dimensional wave equations directly without requiring pre-computed supervised training data. Using vocal tract shape data as input features, we compare the proposed model's predicted $f_0$, glottal volume velocity and sound pressure at the lip for five static vowels to a conventional Runge Kutta/Finite difference approach. With errors of 0.8\% for glottal volume flow and 3.2\% for speech waveforms, the proposed model enables efficient GPU-parallelized simulation without iterative calculations. We conclude that PINO is a promising approach for fast analysis of speech.
\end{abstract}

\section{\label{Sec_Intro} Introduction}

Speech production analysis based on a coupled physical model of the vocal folds and the vocal tract plays an important role in investigating vocal-fold dynamics and diagnosing voice disorders \cite{mcgowan2012source,falk20213d}. Such coupled models enable, for example, estimation of vocal-fold states from speech signals \cite{drioli2011voice,zhao2020speech} and simulation of voice changes associated with surgical treatments for voice disorders \cite{mittal2011toward, rios2020computational}. However, conventional solvers such as the finite element method can be computationally expensive depending on the problem scale \cite{arnela2022tuned}. Moreover, since conventional solvers are typically designed for forward simulations, inverse analysis often requires developing specialized algorithms \cite{zhao2023deriving}.

Neural operators \cite{azizzadenesheli2024neural} offer a promising direction to address these limitations. A neural operator is a machine-learning model that learns mappings from functions to functions, and in numerical simulation it serves as a surrogate that can perform fast, mesh-free simulations across a range of conditions. Applications of neural operators in acoustics have been increasing, including real-time estimation of room impulse responses \cite{borrel2024sound}. In addition, physics-informed neural operators (PINOs) \cite{goswami2023physics, li2024physics, rosofsky2023applications,wang2021learning}, which integrate neural operators with physics-informed neural networks (PINNs) \cite{raissi2019physics} by incorporating governing-equation constraints into the loss function, have been proposed to eliminate the need for supervised training data. In acoustics, PINOs are increasingly being applied to fundamental acoustic wave-scattering problems \cite{nair2025physics}.

Despite these advances in machine-learning-based numerical solvers, their application to speech remains limited. We reported a PINN-based vocal-tract analysis \cite{yokota2024synthesis} and coupled vocal-fold–vocal-tract simulation \cite{yokota2026physics}. Lu and Reiss performed vocal-tract analysis by training PINNs on the Webster equation \cite{lu2026learning}. These remain among the few examples to date. These approaches also benefit from a key advantage of PINNs: the same network used for forward simulation can be used for inverse analysis, making it promising for a wide range of speech-related inverse problems. However, these approaches rely on classical PINNs, which require time-consuming retraining whenever analysis conditions change. This motivates using PINOs, which enable fast simulations across varying analysis conditions.

In this paper, we propose the first PINO framework for speech production analysis that outputs vocal-fold vibration waveforms, glottal flow rate, and speech waveforms at high speed for multiple vocal-tract shapes. Our model is based on Deep Operator Network (DeepONet) \cite{lu2021learning}, a type of neural operator, and employs the two-mass model of Ishizaka and Flanagan \cite{ishizaka1972synthesis} for the vocal folds and a one-dimensional vocal-tract model. Similar to PINNs, we incorporate governing-equation-based loss functions to enable training without supervised data. This approach is referred to as physics-informed DeepONet (PI-DeepONet) \cite{wang2021learning} and has already been applied in the field of musical instrument acoustics \cite{luan2025physicsbow}. By taking vocal-tract shape as an input feature, we demonstrate fast generation of simulation outputs conditioned on different vocal-tract geometries. The validity is verified through comparisons with a conventional numerical solver. The contributions of this work are as follows:
\begin{enumerate}
\item A novel \textit{self-supervised}, \textit{physics-informed} deep learning approach for coupled vocal-fold–vocal-tract simulation across multiple vocal-tract shapes using PINO.
\item The first demonstration of high-speed inference for speech production simulation using a PINO.
\end{enumerate}
\begin{figure*}[t]
\centering
\includegraphics[width={13cm}]{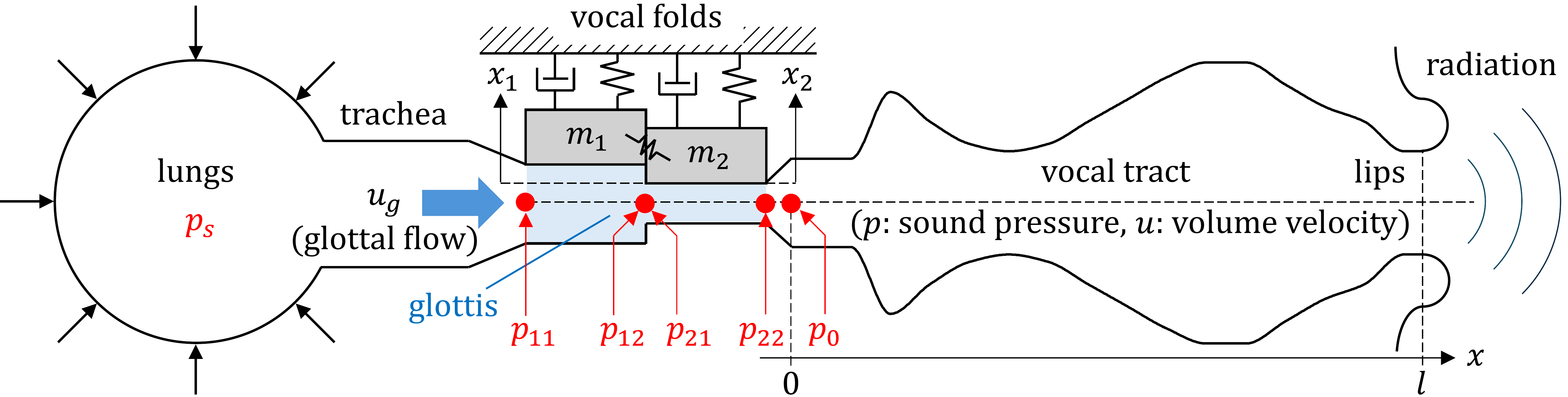}
\caption{\label{Fig_PhysicalModel}{Physical model of speech production used in this study. The vocal folds are modeled using the Ishizaka-Flanagan two-mass model \cite{ishizaka1972synthesis}, and the vocal tract is modeled using a one-dimensional vocal-tract model.}}
\end{figure*}

The remainder of this paper is organized as follows. Section \ref{Sec_PhysicalModel} describes the physical model of speech production. Section \ref{Sec_PINO} presents the proposed PINO framework. Section \ref{Sec_Validation} reports forward-simulation results and demonstrates high-speed output for different vocal-tract shapes. Section \ref{Sec_Conclusion} concludes the paper.

\section{\label{Sec_PhysicalModel} Physical Model of Speech Production}
\subsection{Two-mass Model of Vocal Folds}
In this study, we adopt the two-mass model proposed by Ishizaka and Flanagan \cite{ishizaka1972synthesis} as the model of the vocal folds. As illustrated in Fig.~\ref{Fig_PhysicalModel}, the vocal folds are modeled as a structure with two masses, and the glottal flow is assumed to satisfy Bernoulli’s principle \cite{jiang2002chaotic}. Let $u_g$ denote the glottal volume flow; then, depending on the glottal configuration, the pressures at the points shown in Fig.~\ref{Fig_PhysicalModel} are given by
\begin{align}
p_{11} &= p_s - S_c(x_1) u_g^2, \label{Eq_P_1} \\
p_{12} &= p_{11} - R_{v1}(x_1) u_g, \label{Eq_P_2} \\
p_{21} &= p_{12} - S_{12}(x_1,x_2) u_g^2, \label{Eq_P_3} \\
p_{22} &= p_{21} - R_{v2}(x_2) u_g, \label{Eq_P_4} \\
p_0 &= p_{22} - S_e(x_2) u_g^2. \label{Eq_P_5}
\end{align}
Here, $x_1$ and $x_2$ are the vocal-fold displacements; $p_s$ is the subglottal pressure; $S_c$ represents the inlet contraction at the glottal entrance; $R_{v1}$ and $R_{v2}$ represent viscous losses along the wall surfaces; and $S_{12}$ is a coefficient that accounts for pressure variation due to changes in the glottal configuration. $S_e$ is a coefficient representing pressure recovery due to flow separation and reattachment at the glottal exit. From Eqs. (\ref{Eq_P_1})-(\ref{Eq_P_5}), $u_g$ is obtained as
\begin{equation}
    u_g = \frac{-\beta + \sqrt{\beta^2-4 \alpha \gamma}}{\alpha}.
    \label{Eq_ug}
\end{equation}
Here, $\alpha$, $\beta$, and $\gamma$ are calculated from $S$, $R$, $p_0$, and $p_s$ in Eqs.~(\ref{Eq_P_1})-(\ref{Eq_P_5}). In Eq. (\ref{Eq_ug}), $u_g$ is a function of the vocal-fold displacements $x_1$ and $x_2$.

The dynamics of the displacements $x_1$ and $x_2$ of each mass from its equilibrium position are described by
\begin{align}
    m_1 \ddot{x}_1 + c_1\dot{x}_1 + s_1 + k_c (x_1-x_2) &= f_1, \label{Eq_VF1}\\
    m_2 \ddot{x}_2 + c_2\dot{x}_2 + s_2 + k_c (x_2-x_1) &= f_2. \label{Eq_VF2}
\end{align}
Here, $m_i$, $c_i$, and $k_c$ denote the vocal-fold mass, damping, and coupling spring constant, respectively, $s_i$ represents the stiffness-related force, and $f_i$ is the force acting on the vocal folds due to the glottal pressure.

\subsection{One-dimensional Model of Vocal Tract}
In this study, the vocal tract is modeled using a one-dimensional model \cite{yokota2024synthesis,flanagan2013speech}.
Let $p$ and $u$ denote the acoustic pressure and volume velocity in the vocal tract, respectively. Acoustic wave propagation in the vocal tract is governed by
\begin{align}
\frac{\partial u}{\partial x} &= -Gp - \frac{A}{K}\frac{\partial p}{\partial t}, \label{WaveEq_Time1}\\
\frac{\partial p}{\partial x} &= -Ru - \frac{\rho}{A}\frac{\partial u}{\partial t},
\label{WaveEq_Time2}
\end{align}
where $A$ is the cross-sectional area of the vocal tract, $\rho$ is the air density, $K$ is the bulk modulus, and $R$ and $G$ are energy loss parameters. Acoustic radiation at the lips is modeled following \cite{ishizaka1972synthesis,thibault2020time} and is given by the following equation:
\begin{equation}
L_r \dfrac{du_l}{dt} = p_l + \dfrac{L_r}{R_r} \dfrac{dp_l}{dt}, \label{Radiation}
\end{equation}
where $L_r$ and $R_r$ are physical parameters related to radiation, and $p_l$ and $u_l$ denote the acoustic pressure and volume velocity at the opening ($x=l$), respectively.

\section{\label{Sec_PINO} PINO for Speech Production}
\subsection{Network Architecture}
\begin{figure*}
\centering
\includegraphics[width={15cm}]{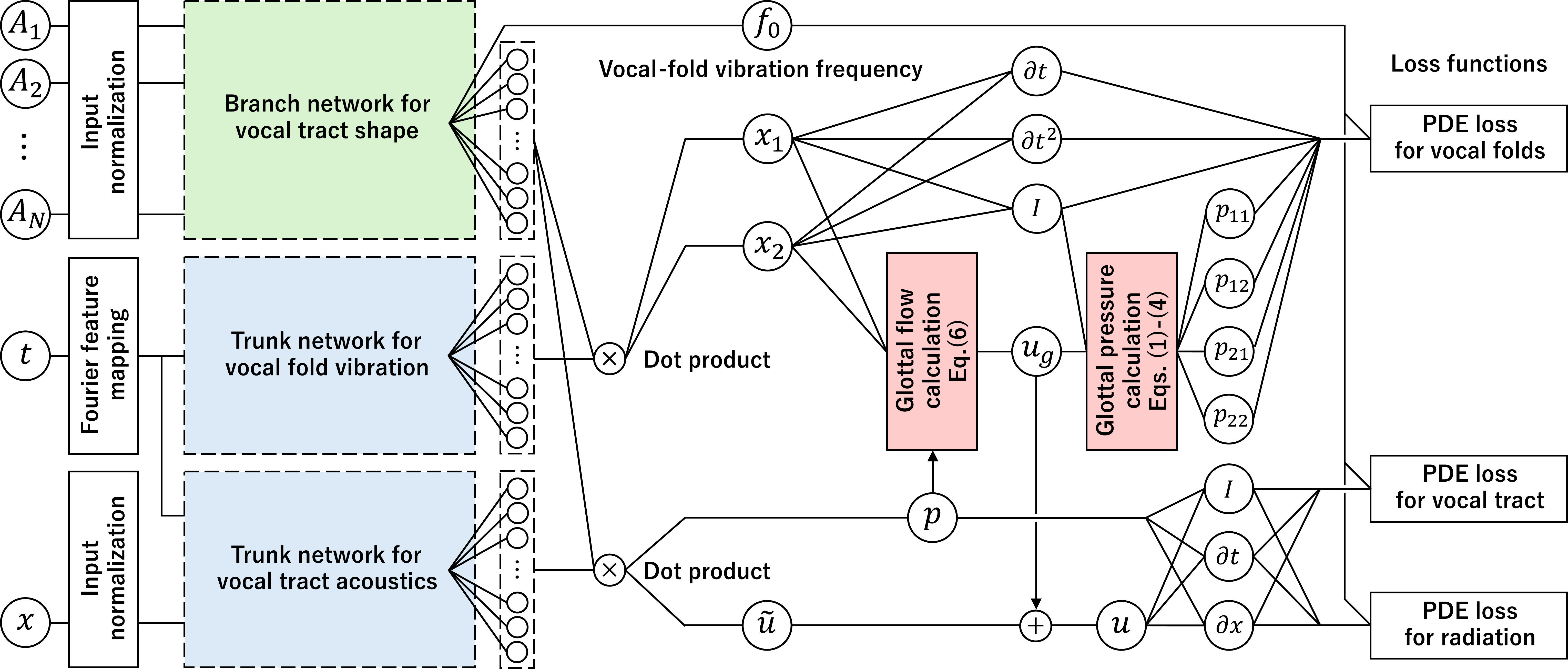}
\caption{\label{Fig_PINO}{Proposed PINO for speech production. The branch network predicts the vocal-fold vibration frequency $f_0$ conditioned on the vocal-tract shape, enabling steady-state analysis. The loss calculation follows our previously proposed procedure \cite{yokota2026physics}.}}
\end{figure*}

Figure~\ref{Fig_PINO} shows the network architecture of the proposed PINO. This architecture is based on PI-DeepONet \cite{lu2021learning}, a type of neural operator, and consists of a branch network that takes the normalized vocal-tract shape as input and a trunk network that takes the collocation point ($x$ and $t$) as input. For both networks, we employ a serial stack of fully connected (FC) blocks used in ResoNet \cite{yokota2024physics}, and use the Snake activation function \cite{ziyin2020neural}.

In the branch network, the input is the vocal-tract cross-sectional areas $A_1, A_2, \ldots, A_N$ sampled at uniform intervals along the $x$-axis. This network outputs the branch features and the steady-state vocal-fold vibration frequency $f_0$ for the given shape. By taking the inner product between the branch and the trunk features, the model outputs the vocal-fold displacements $x_1$ and $x_2$, the acoustic pressure $p$ in the vocal tract, and $\tilde{u}$, which is used to compute the volume velocity. $\tilde{u}$ is converted to the volume velocity $u$ via the following hard constraint for vocal-fold-vocal-tract coupling \cite{yokota2026physics}:
\begin{equation}
    u = \tilde{u} \sin{\frac{\pi}{2l}x} + u_g \left( 1 - \sin{\frac{\pi}{2l}x} \right).
\label{Eq_CouplingMethod}
\end{equation}
Because this constraint guarantees $u=u_g$ at $x=0$, the coupling between the vocal folds and the vocal tract is satisfied as a hard constraint, without requiring an additional loss function. Note that $u_g$ is constrained to be positive in the network using our previously proposed method \cite{yokota2026physics}, consistent with the assumption of the two-mass model.

The operations applied to $x_1$, $x_2$, $\tilde{p}$, and $\tilde{u}$ on the right-hand side of Fig.~\ref{Fig_PINO} follow the PINN for speech production we proposed \cite{yokota2026physics}. To enforce the equations of motion for the vocal folds, Eqs. (\ref{Eq_VF1}) and (\ref{Eq_VF2}), we define the following PDE (partial differential equation) losses:
\begin{equation}
    \begin{split}
        \mathcal{L}_{f1} &= \dfrac{1}{N_f} \sum_{i=1}^{N_f} \left\{
        m_1 \ddot{x}_{1i} + c_1\dot{x}_{1i} + s_{1i} \right. \\
        &\left. \quad + k_c (x_{1i}-x_{2i}) - f_{1i} \right\}^2, \label{Eq_Loss_VF1}
    \end{split}
\end{equation}
\begin{equation}
    \begin{split}
        \mathcal{L}_{f2} &= \dfrac{1}{N_f} \sum_{i=1}^{N_f} \left\{
        m_2 \ddot{x}_{2i} + c_2\dot{x}_{2i} + s_{2i} \right. \\
        &\left. \quad  + k_c (x_{2i}-x_{1i}) - f_{2i} \right\}^2, \label{Eq_Loss_VF2}
    \end{split}
\end{equation}
where $i$ is the collocation index and $N_f$ is the number of collocation points. Similarly, to enforce the vocal-tract equations, Eqs. (\ref{WaveEq_Time1}) and (\ref{WaveEq_Time2}), the following PDE losses are defined:
\begin{align}
\mathcal{L}_{t1} &= \dfrac{1}{N_t} \sum_{i=1}^{N_t} \left(
\frac{\partial u_i}{\partial x_i} + G_i p_i + \frac{A_i}{K}\frac{\partial p_i}{\partial t_i}
\right)^2, \label{Eq_Loss_Time1}\\
\mathcal{L}_{t2} &= \dfrac{1}{N_t} \sum_{i=1}^{N_t} \left(
\frac{\partial p_i}{\partial x_i} + R_i u_i + \frac{\rho}{A_i}\frac{\partial u_i}{\partial t_i}
\right)^2. \label{Eq_Loss_Time2}
\end{align}
Finally, to enforce the radiation equation at the lips, Eq. (\ref{Radiation}), the following PDE loss is introduced:
\begin{equation}
\mathcal{L}_r = \dfrac{1}{N_r} \sum_{i=1}^{N_r} \left(
L_r \dfrac{du_{l,i}}{dt_i} - p_{l,i} - \dfrac{L_r}{R_r} \dfrac{dp_{l,i}}{dt_i}
\right)^2. \label{Eq_Loss_Rad}
\end{equation}
The total loss function for the entire network, $\mathcal{L}_{all}$, is given by
\begin{equation}
    \mathcal{L}_{all} = \lambda_f \left( \mathcal{L}_{f1}+\mathcal{L}_{f2} \right) 
    + \lambda_{t1} \mathcal{L}_{t1} + \lambda_{t2} \mathcal{L}_{t2} + \lambda_{r} \mathcal{L}_r,
    \label{Eq_TotalLoss}
\end{equation}
where $\lambda$ denotes the weight coefficients for each loss function.

\subsection{\label{Sec_Steady} Steady-State Analysis for Different Vocal-Tract Shapes}
To mitigate spectral bias \cite{wang2022and}, our method analyzes only one period of the steady-state resonance. Let $t^{*}$ denote the time input normalized to the range $[-1,1]$, and $t^{*}$ is mapped using Fourier features \cite{lu2021physics} as
\begin{equation}
    \bm{t}^{*} = \left[
    \cos{\pi t^{*}},\ \sin{\pi t^{*}}, \cdots, \ \cos{m \pi t^{*}},\ \sin{m \pi t^{*}} \right],
    \label{Eq_FF}
\end{equation}
where $m$ is the number of Fourier features (positive integer). Since the above mapping yields the same value at $t=0$ and $t=T$ ($T$: period), periodicity is imposed as a hard constraint.

Because the vocal-fold vibration frequency $f_0$ varies with the vocal-tract shape, the proposed method outputs $f_0$ from the branch network.
Using this, the time-derivative terms in Eqs. (\ref{Eq_Loss_VF1})-(\ref{Eq_Loss_Rad}) are computed as
\begin{equation}
    \frac{\partial}{\partial t} = 2 f_0 \frac{\partial}{\partial t^*}.
    \label{NormalizeT}
\end{equation}
Eq. (\ref{NormalizeT}) indicates that the time derivative in the original domain is obtained by scaling the derivative with respect to the normalized time $t^*$ (defined over the range $[-1,1]$) by a factor of $2/T$ ($=2f_0$). By computing the PDE loss using Eq. (\ref{NormalizeT}), the vocal-fold vibration frequency for each vocal-tract shape can be learned without changing the input collocation points.

\subsection{Implementation}
The proposed method was implemented in MATLAB. Training and inference were performed on a PC equipped with an Intel Core Ultra 9 285K CPU and an NVIDIA RTX PRO 6000 Blackwell Workstation Edition GPU.

\section{\label{Sec_Validation} Validation of the Proposed Method}
\subsection{Analysis Conditions}
\begin{figure}[t]
\centering
\includegraphics[width={7cm}]{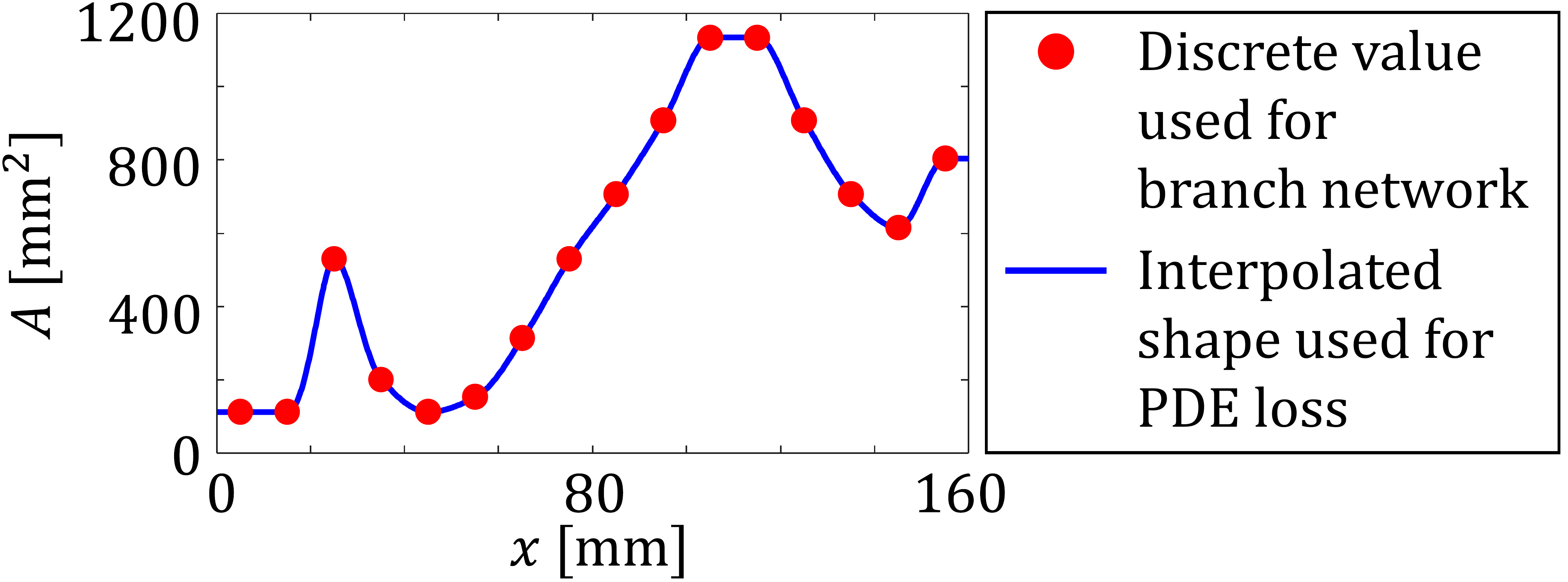}
\caption{\label{Fig_VT_Shape} Vocal-tract cross-sectional area function for vowel /a/. Red markers indicate the discrete values used as input features to the branch network.}
\end{figure}

The validity of the proposed method is evaluated through speech production simulations for multiple vowel configurations. We use five vowel-area functions for /a/, /i/, /u/, /e/, and /o/ reported by Arai~\cite{arai2007education}, and train a single network on all five shapes. The vocal-tract shapes reported by Arai are given as tract diameters sampled every 1~cm along the $x$-axis; as an example, the corresponding cross-sectional areas for /a/ are shown as red markers in Fig.~\ref{Fig_VT_Shape}.
As the input features to the branch network in Fig.~\ref{Fig_PINO}, we use these 16 sampled cross-sectional areas. To compute Eqs.~(\ref{WaveEq_Time1}) and (\ref{WaveEq_Time2}), the discrete samples (red markers) are interpolated using the piecewise cubic Hermite interpolating polynomial (PCHIP) method \cite{PCHIP1,PCHIP2}. The interpolated vocal-tract shape is shown as the blue curve in Fig.~\ref{Fig_VT_Shape}. The physical parameters are identical to our previous values~\cite{yokota2026physics}.

The number of FC blocks~\cite{yokota2024physics} was set to 3 for the branch network, 3 for the vocal-fold trunk network, and 5 for the vocal-tract trunk network. Each fully connected layer had 200 nodes. We used 41,000 collocation points for each vowel, resulting in 205,000 collocation points in total for the five vowels. Network parameters were optimized using the Adam optimizer~\cite{Adam}.

For performance comparison, we also conducted simulations using conventional solvers: a fourth-order Runge-Kutta (RK4) method for the vocal-fold model and a finite-difference method (FDM) implementation for the vocal-tract model. The step sizes of the conventional solver were $\Delta x = 1.0 \times 10^{-4}$~m and $\Delta t = 5.88 \times 10^{-8}$~s.

\subsection{Simulation Results}
\begin{figure}
\centering
\includegraphics[width={6cm}]{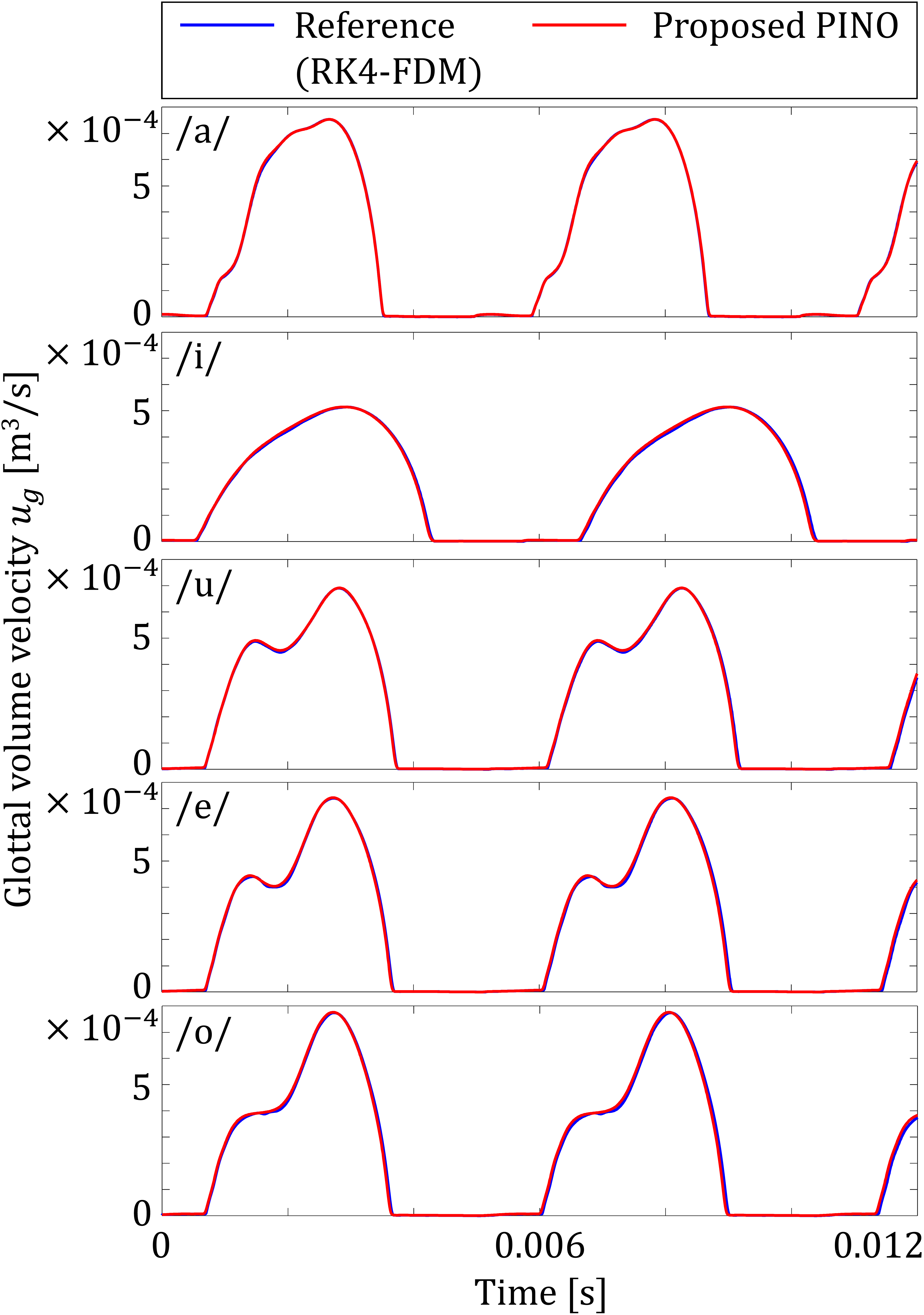}
\caption{\label{Fig_Ug} Simulated glottal volume flow $u_g$. As described in Sec. \ref{Sec_Steady}, the proposed method analyzes only one period of the steady state; however, the same waveform is repeatedly displayed to confirm periodicity.}
\end{figure}

\begin{figure}[t]
\centering
\includegraphics[width={6cm}]{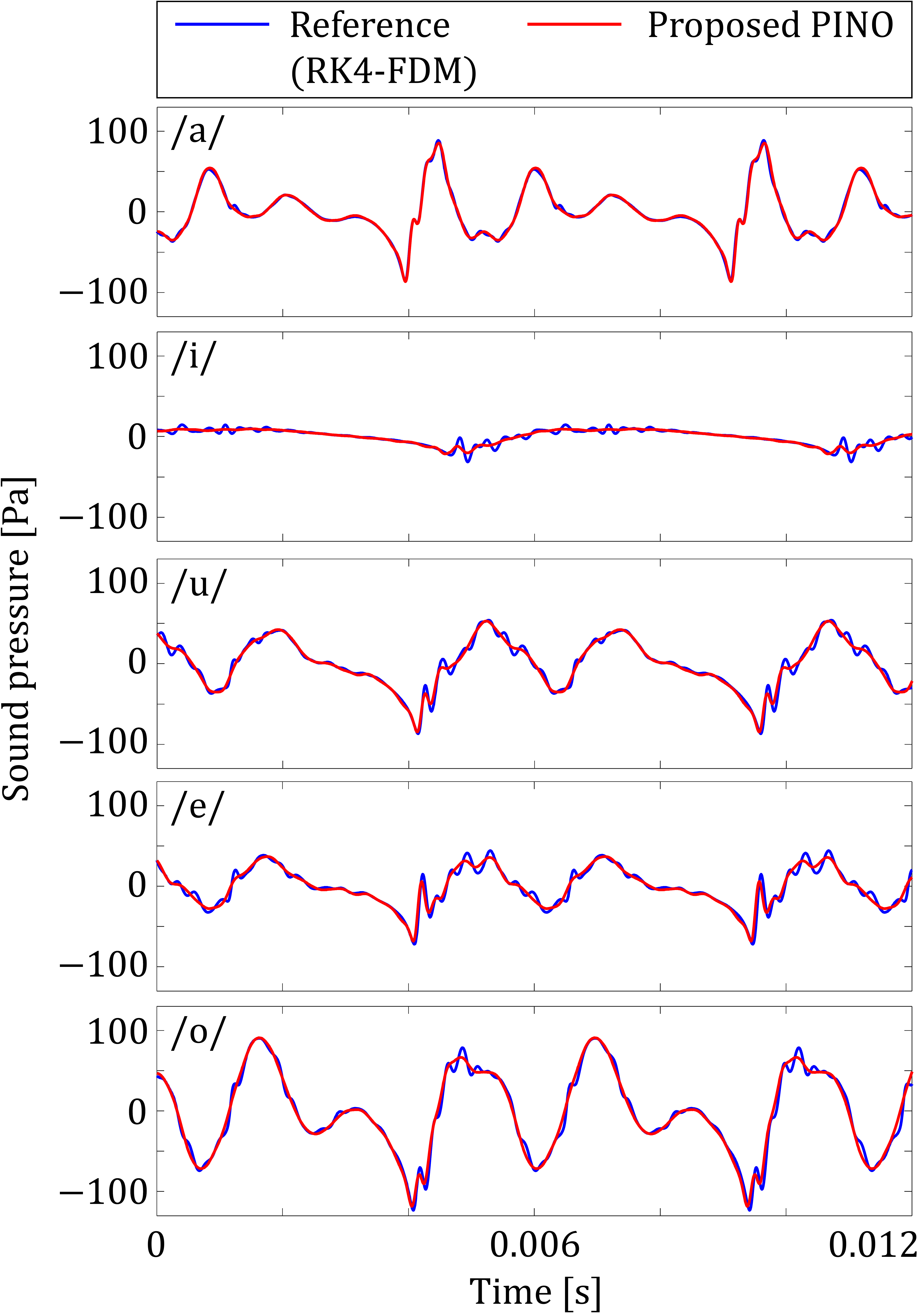}
\caption{\label{Fig_P_lips} Simulated sound pressure at the lips $p_l$.}
\end{figure}

Table~\ref{Table_Freq} summarizes the vocal-fold vibration frequency $f_0$ estimated by the branch network after 100,000 training epochs. The estimated $f_0$ values closely match the reference values obtained with the conventional RK4-FDM solver. Figure~\ref{Fig_Ug} shows the glottal volume flow rate $u_g$ for each vowel, and Fig.~\ref{Fig_P_lips} shows the acoustic pressure waveform at the lips. For each vocal-tract shape, the proposed method produces waveforms that agree well with the reference results.
\begin{table}
  \centering
  \caption{Estimated vocal-fold vibration frequency $f_0$.}
  \setlength{\tabcolsep}{6pt}
  \begin{tabular}{l|c|c|c|c|c}
    \toprule
    & /a/ & /i/ & /u/ & /e/ & /o/ \\
    \midrule
    Reference [Hz] & 193.3 & 164.0 & 183.6 & 186.0 & 187.1 \\
    Estimated [Hz] & 193.1 & 164.3 & 183.9 & 186.3 & 187.5\\
    Difference [\%] & 0.105 & 0.196 & 0.194 & 0.185 & 0.214\\
    \bottomrule
  \end{tabular}
  \label{Table_Freq}
\end{table}
\begin{table}
  \centering
  \caption{Range-normalized RMSE showing the differences between the PINO and the conventional solver in Figs. \ref{Fig_Ug} and \ref{Fig_P_lips}. Lower values indicate higher analysis accuracy of the PINO.}
  \setlength{\tabcolsep}{6pt}
  \begin{tabular}{l|c|c|c|c|c}
    \toprule
    & /a/ & /i/ & /u/ & /e/ & /o/ \\
    \midrule
    $u_g$ (Fig. \ref{Fig_Ug}) [\%] & 0.36 & 1.00 & 0.50 & 0.91 & 1.23 \\
    $p_l$ (Fig. \ref{Fig_P_lips}) [\%] & 1.01 & 5.98 & 2.66 & 3.97 & 2.47\\
    \bottomrule
  \end{tabular}
  \label{Table_RMSE}
\end{table}

For quantitative comparison, Table \ref{Table_RMSE} presents the range-normalized root-mean-square error (RMSE) between the predicted waveforms and the references shown in Figs. \ref{Fig_Ug} and \ref{Fig_P_lips}. The sound pressure exhibits a relatively higher error compared to the glottal volume flow; this is likely because the sound pressure contains higher formant frequencies, which are challenging for the PINO to approximate due to spectral bias \cite{wang2022and}. To address this challenge, input encoding methods such as multi-scale Fourier features \cite{wang2021eigenvector} are considered a promising direction.

Although the training for the five vowels required approximately 80 hours, the average inference time per vowel was 0.0389 s. Unlike conventional solvers that require iterative calculations at each time step, the PINO enables efficient simulation through parallel computing on GPUs. Consequently, it is expected to further evolve as a high-speed speech production simulation method in the future.

\section{\label{Sec_Conclusion} Conclusion}
In this study, we developed a PINO-based speech production analysis model for fast speech production simulation across diverse vocal-tract shapes. The proposed model takes the vocal-tract shape as an input feature and outputs the corresponding vocal-fold vibration frequency, enabling computation of steady-state periodic solutions for multiple vocal-tract geometries. Results obtained by training on five vowel vocal-tract shapes demonstrate that the proposed method enables fast inference of simulation outputs.

Although the current analysis focuses only on inference for trained vocal-tract shapes and is limited to steady-state conditions, future research will focus on improving generalization performance for diverse vocal-tract shapes, addressing non-steady-state conditions, enhancing training efficiency, expanding to three-dimensional geometries, and investigating applications to consonant synthesis.

\section{Acknowledgments}
This work was supported by JSPS KAKENHI Grant Number JP25K03137, JSPS Program for Forming Japan's Peak Research Universities (J-PEAKS) Grant Number JPJS00420240017 and the Ono Charitable Trust for Acoustics.

\section{Generative AI Use Disclosure}

There is no generative AI use to disclose for this study.

\bibliographystyle{IEEEtran}
\bibliography{mybib}

\end{document}